\documentclass[twocolumn]{revtex4}
\usepackage{graphicx}
\begin{document}
\title {The most robust entangled state of light}
\author{S.J. van Enk$^{1,2}$ and O.~Hirota$^3$}
\affiliation{$^1$Bell Labs, Lucent Technologies\\
600-700 Mountain Ave,
Murray Hill, NJ 07974\\
$^2$Institute for Quantum Information, California Institute of Technology\\
Pasadena, CA 91125\\
$^3$Research Center for Quantum Communications, \\
Tamagawa University,
Tokyo, Japan 
}

\date{\today}

\begin{abstract}
We study how photon absorption losses degrade the bipartite
entanglement of entangled
states of light. We consider two questions:
(i) what state contains the smallest average number of photons
given a fixed
amount of entanglement? and 
(ii) what state is the most robust against photon absorption?
We explain why the two-mode squeezed state is the answer to the first question
but not quite to the second question.
\end{abstract}

\maketitle
\section{Introduction}
For long-distance quantum communication the most
important decoherence mechanism is due to photon absorption.
It would be nice to know what type of entangled states of light is most robust,
i.e. preserves its entanglement best,
against photon absorption losses. 
Intuitively one might think that states containing fewer 
photons will be more robust. After all, both loss rates
and decoherence rates tend to increase 
with the number of photons. For instance, 
a cavity filled with exactly $N$ photons 
loses photons at a rate proportional to $N$. Similarly, it is well known 
that, in general, 
the more macroscopic a state is the more sensitive it is to noise.
For example, the decoherence rate of
Schr\"odinger cat states of the (unnormalized) form 
\begin{equation}
|\alpha\rangle\pm|-\alpha\rangle,
\end{equation}
with $|\alpha\rangle$ a coherent state,
increases with the amplitude $|\alpha|$. This  
phenomenon was demonstrated experimentally 
in the context of microwave cavity-QED \cite{haroche}
and for 
motional states of ions in an ion trap \cite{myatt}. 
Also entangled coherent states of the (unnormalized) form
\begin{equation}
|\alpha\rangle|\alpha\rangle
-|-\alpha\rangle|-\alpha\rangle,
\end{equation}
decohere
faster with increasing $|\alpha|$.
This fact was discussed in several theoretical papers: Ref.~\cite{hirota}
uses an operational
measure and
shows that the teleportation fidelity decreases monotonically
with $|\alpha|$, and another paper \cite{lixu} demonstrates that
the entanglement
of formation decreases monotomically with $|\alpha|$ as well. 

If the intuition that a smaller number of photons makes a state more robust
were true, then the most robust entangled state of 
light  
would be the two-mode squeezed state. That state contains
 the smallest average number of photons given a fixed amount of entanglement, 
as we will show in Section II.
We will see, however, that the intuitive answer is not quite correct
and we will explain why this is so in Section III.
The calculations in Section III proceed as follows: we consider 
large classes of pure states (both Gaussian and non-Gaussian states)
with a fixed amount of entanglement.
Then we calculate the entanglement that 
remains after those states have decohered
by being subjected to photon
absorption. That remaining entanglement is 
considered to be a measure of the robustness of 
the entangled state against photon absorption. The robustness is investigated
as a function of the average number of photons in the initial pure state.
We use two definitions of entanglement, the entanglement of formation
and the negativity. In the former case we restrict our attention to states
for which the entanglement of formation can be calculated analytically,
in the latter case there is no restriction on the types of states 
considered.

We end this Introduction by noting that
the
entanglement of a state of the electromagnetic field
is completely determined by its expansion in Fock 
(photon number) states \cite{pz}.
Consequently,
the effects of
photon absorption losses on the entanglement of a state
are likewise completely determined by the form of 
the state in the Fock basis.
On the other hand, the effects of
other decoherence mechanisms arising from, say, polarization diffusion or
phase diffusion, 
depend on the physical implementation of the state. 
For example, if one uses two modes that are different only in their
polarization degree of freedom, then
polarization diffusion mixes the two modes. But if the two modes are
spatially separated then no such mixing occurs.
The type of decoherence discussed in the present paper is thus
the most universal type of decoherence of entangled states of light, in the
sense that it is
independent of the precise physical implementation of the state.

\section{Photons and entanglement}
Let us consider a bipartite system
consisting of two spatially separated electromagnetic field modes $A$ and $B$. 
The entanglement in a pure state
on $A$ and $B$ is given by the von Neumann entropy of the partial
density matrix of either system $A$ or $B$, 
where the pure state can be written in the photon-number basis as
\begin{equation}\label{AB}
|\Psi\rangle_{A,B}=\sum_{n,m=0}^\infty\alpha_{nm}|n\rangle_A|m\rangle_B.
\end{equation}
The problem of finding the state with the smallest average 
number of photons for a fixed amount of entanglement 
is mathematically
equivalent to a well-known problem from thermodynamics: 
finding the state with the largest entropy given a fixed energy,
which is, almost obviously, the same as finding the state with the 
smallest energy given a fixed entropy. 
Indeed, energy is proportional to the average number of photons, 
and the entropy is just the entanglement. The answer 
to the thermodynamics question is of course the thermal state.
That is, the state
\begin{equation}
\rho=(1-\lambda)\sum_n \lambda^n |n\rangle \langle n|
\end{equation}
maximizes the entropy for fixed average energy, 
where $\lambda=\exp(-\hbar\omega/kT)$
if $n\hbar\omega$ is the energy of the state $|n\rangle$ and $T$ the 
temperature.
But that mixed state arises indeed from a two-mode squeezed state
of the form
\begin{equation}\label{tm}
\sqrt{1-|\gamma|^2}\sum_n \gamma^n |n\rangle |n\rangle,
\end{equation}
for any $\gamma$ for which $|\gamma|^2=\lambda$.
In particular, the entanglement in this state is
\begin{equation}
E=(\frac{\bar{n}}{2}+1)\log_2(\bar{n}+2)-\bar{n}\log_2\bar{n}-1,
\end{equation}
in terms of the average number of photons
\begin{equation}
\bar{n}=\frac{2\lambda}{1-\lambda}=\frac{2|\gamma|^2}{1-|\gamma|^2}.
\end{equation}
For instance, for $\bar{n}=1$ 
one has more than one unit of entanglement,
as $E=3\log_2(3)/2-1\approx 1.3774$.
Conversely, it takes less than one photon 
($0.5876$ of a photon to be more precise)
to have one ebit of entanglement between two modes. 
That might seem surprising
as one might have expected the state $|0\rangle|1\rangle+|1\rangle|0\rangle$,
a state with one photon and one ebit of entanglement, to 
maximize the entanglement of one photon in two modes.

If one allows an arbitrarily large number of modes for both
parties, it is straightforward 
to get an arbitrarily large amount of bipartite entanglement. For example,
consider states with two photons: with $M$ modes for each party,
the bipartite state
$|1000\ldots1000\ldots\rangle+
|0100\ldots0100\ldots\rangle+|0010\ldots0010\ldots\rangle+\ldots$ has
exactly $\log_2(M)$ ebits of entanglement. 
But the state that maximizes the amount of entanglement in the case of 
$M$ modes for both parties
for two photons contains more than twice as much entanglement
and is simply a tensor product of $M$
two-mode squeezed states each with average photon number $2/M$,
with a total entanglement of $2\log_2(M)+1+{\cal O}(1/M)$ for large $M$.
\section{Entanglement of decohered states}
In this Section we will consider the most robust bipartite state
of just {\em two} modes. This should reveal most properties
of multi-mode entangled states under the assumption that
all modes decohere independently and identically. In particular,
we assume the following model for photon absorption losses:
For any coherent state $|\alpha\rangle_A$ of a mode $A$, photon absorption
is governed by the process
\begin{equation}\label{model}
|\alpha\rangle_A|0\rangle_E\mapsto
|\sqrt{\eta}\alpha\rangle_A|\sqrt{1-\eta}\alpha\rangle_E,
\end{equation}
where $E$ denotes the environment, which is assumed to be unobservable and 
hence must be traced out.
The parameter 
$\eta$ gives the fraction of photons
in mode $A$ that survives the photon absorption process, 
with $\eta=1$ thus corresponding to
a noiseless channel.
Since coherent states form an overcomplete set of states,
the model (\ref{model}) fully specifies decoherence due to photon
absorption. We assume that all modes decohere in the same way, according to
(\ref{model}), with the same
parameter $\eta$ for each mode and a different environment for each mode.

A pure state of the form (\ref{AB}) is turned into a mixture by the photon 
absorption process (\ref{model}). Calculating the 
entanglement of a general
mixed state is no trivial problem, as is well known.
We follow two paths here in order to find the entanglement that remains after
photon absorption. 

In the first subsection we consider states
for which we can analytically
calculate the entanglement of formation for the mixed state
that arises from photon absorption.
This first of all includes all states
in which each of the two modes $A$ and $B$ contain at most one photon. 
In that case the Hilbert space of each mode is two-dimensional, 
also after photon absorption, and we can apply 
the Wootters formula \cite{wootters}. Secondly, 
we can consider the two-mode
squeezed state, as it is a Gaussian state and remains Gaussian
after photon absorption,
so that we can apply
the results of \cite{giedke}.  
In the second subsection we consider arbitrary states and calculate 
(numerically)
the negativity \cite{vidal1,vidal2}. Here too, analytical calculations
become tedious or impossible depending on the dimension of the 
matrix whose eigenvalues have to be calculated.

In both subsections
we consider states with a fixed amount
of entanglement and wish to find
the states that preserve their entanglement best 
under photon absorption.
\subsection{Entanglement of formation}
We first consider states of the general form
\begin{equation}\label{form1}
\sqrt{p}|\phi\rangle_A|\varphi\rangle_B
+\sqrt{1-p}|\phi^\perp\rangle_A|\varphi^{\perp}\rangle_B,
\end{equation}
where the parameter $0\leq p\leq 1/2$ 
fully determines the entanglement of 
formation of the state 
\begin{equation}
E=-p\log_2(p)-(1-p)\log_2(1-p),
\end{equation}
and where
\begin{eqnarray}
|\phi\rangle_A&=&\cos\alpha |0\rangle+\sin\alpha|1\rangle
\nonumber\\
|\varphi\rangle_B&=&\cos\beta |0\rangle+\sin\beta|1\rangle
\nonumber\\
|\phi^\perp\rangle_A&=&-\sin\alpha |0\rangle+
\cos\alpha|1\rangle
\nonumber\\
|\varphi^\perp\rangle_B&=&-\sin\beta |0\rangle
+\cos\beta|1\rangle
\end{eqnarray}
form arbitrary orthonormal
bases in the subspaces spanned by the zero- and one-photon
states in modes $A$ and $B$. 
We then fix $p$ and $\eta$ and plot the entanglement
of all states of the form (\ref{form1}) after they have lost
a fraction $1-\eta$ of their photons
by varying the angles $\alpha$
and $\beta$ independently.
We plot the result
as a function of the average number of photons $\bar{n}$
in the original
undecohered pure state (\ref{form1}),
\begin{equation}
\bar{n}=p(\cos^2\alpha+\cos^2\beta)+(1-p)(\sin^2\alpha+\sin^2\beta).
\end{equation}
Results for specific parameters are plotted in Fig.~1.
\begin{figure}
\includegraphics[scale=0.4]{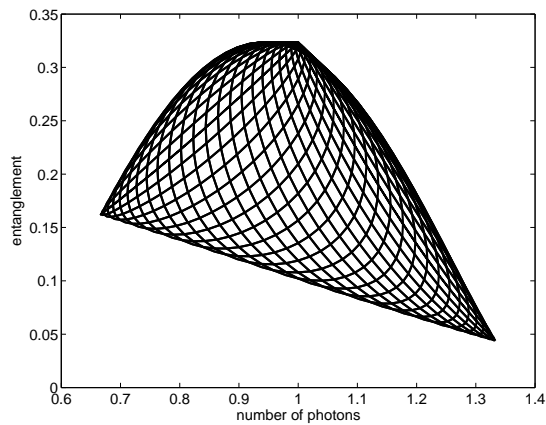}
\caption{Entanglement of formation 
as a function of the average number of
photons (in both modes combined) 
for all bipartite states with at most one photon
in each mode.
}
\end{figure}
In particular, we choose $\eta=0.5$ and fix $p=1/3$, so that the entanglement
of formation of the pure state is $E\approx 0.9183$.
The optimum state is found to be 
\begin{equation}\label{op}
\sqrt{p}|01\rangle-\sqrt{1-p}|10\rangle,
\end{equation}
which contains one photon and 
for which the entanglement left after decoherence is $E\approx 0.3236$.
Now clearly, this optimum is not obtained for the 
smallest average number of photons, which is 2/3 for the states (\ref{form1})
considered.
That minimum average number of photons is attained for the state
\begin{equation}\label{minp}
\sqrt{p}|11\rangle-\sqrt{1-p}|00\rangle,
\end{equation}
but its entanglement after decoherence is only $E\approx 0.1622$.
(One notices the similarity of the state (\ref{minp}) and the
two-mode squeezed state.)
These two facts clearly refute the intuition that a lower number of photons
should lead to less decoherence.
On the other hand, it is true that the state with the largest 
amount of photons, 4/3, decoheres the most. 
That is, in the state
\begin{equation}\label{maxp}
\sqrt{1-p}|11\rangle+\sqrt{p}|00\rangle
\end{equation}
the amount of entanglement surviving the 50\% photon absorption
is $E\approx 0.0438$.
The minimum entanglement after decoherence for a fixed average number
of photons is in fact obtained for the family of states
for which $\alpha=\beta$.

In order to explain
the preceding results, let us give an 
intuitive idea for why the state
\begin{equation}\label{1100}
\sqrt{1/2}|11\rangle+\sqrt{1/2}|00\rangle
\end{equation}
is less robust (when using the entanglement of formation as
entanglement measure) than the state
\begin{equation}\label{1001}
\sqrt{1/2}|10\rangle+\sqrt{1/2}|01\rangle,
\end{equation}
although both states contain 1 photon on average.
For concreteness, again assume we lose 50\% of the photons.
Suppose we would be able to
measure the number of photons in the two environments 
into which the two modes decohere (with a perfect photon detector). 
In that case, if we don't find any photons
we know we have an entangled state of the two modes, 
whereas if we do find a photon in the environment,
we have no entanglement left between the two modes.
In the case of the state (\ref{1100}) we have a probability of
$(1+\eta^2)/2=5/8$ to find no photons in the environment, but the
state of the two modes is collapsed onto the (unnormalized)
state
\begin{equation}\label{1100d}
\sqrt{1/2}|11\rangle+\sqrt{1/2}\eta|00\rangle
\end{equation}
with an entanglement of $E\approx 0.72$ for $\eta=1/2$.
On the other hand, for the state (\ref{1001})
we have a slightly smaller chance of $\eta=1/2$ to find no
photons in the environment, but the state of the two modes then collapses
back onto (\ref{1001}), with its full entanglement of 1 ebit.
Thus on average we indeed retain more entanglement from
state (\ref{1001}) (namely 0.5 ebits) 
than from (\ref{1100}) (namely, 0.45 ebits) after decoherence,
making (\ref{1001}) the more robust state.
This then also indicates why the family of states (\ref{op})
are more robust than either of the families of states
(\ref{minp}) and (\ref{maxp}).
 
A second type of states for which the entanglement of formation
after decoherence can be calculated is the set of entangled
coherent states. The reduced density matrix of those states 
has rank two,  the support given by the 
two-dimensional Hilbert space spanned by 
two different coherent states, which we choose here as 
$|\alpha\rangle$ and $|-\alpha\rangle$. After decoherence the
relevant Hilbert space remains two-dimensional,
spanned by coherent states $|\pm\sqrt{\eta}\alpha\rangle$.
We can distinguish three types of states.
The first two are both symmetric under the interchange 
of the two modes $A$ and $B$,
\begin{equation}\label{ech1}
\sqrt{p}|+\rangle_A|+\rangle_B
+\sqrt{1-p}\exp(i\phi)|-\rangle_A|-\rangle_B,
\end{equation}
with $0\leq p\leq 1/2$, and
\begin{equation}\label{ech2}
\sqrt{1-p}|+\rangle_A|+\rangle_B
+\sqrt{p}\exp(i\phi)|-\rangle_A|-\rangle_B.
\end{equation}
The third type is of the general form
\begin{equation}\label{ech3}
\sqrt{p}|+\rangle_A|-\rangle_B
+\sqrt{1-p}\exp(i\phi)|-\rangle_A|+\rangle_B.
\end{equation}
Here the $|\pm\rangle$ states are defined by
\begin{eqnarray}
|\pm\rangle&=&(|\alpha\rangle\pm|-\alpha\rangle)/\sqrt{N_{\pm}}\nonumber\\
N_{\pm}&=&2\pm2\exp(-2|\alpha|^2),
\end{eqnarray}
which form an 
orthonormal basis for the relevant Hilbert space.
Just as before we choose $p=1/3$ and $\eta=1/2$.
The entanglement left after decoherence
is plotted
in Fig.~2 for various values of the phase $\phi$.
One clearly identifies three groups of curves. They correspond to
the states (\ref{ech1}), (\ref{ech2}), and (\ref{ech3}).
In particular, the states with the largest amount of entanglement
correspond to (\ref{ech3}), 
and the states with the smallest and largest numbers
of photons, respectively, to (\ref{ech1}) and (\ref{ech2}). 
Within each group the states have the property
that fewer photons lead to less decoherence. On the other hand, the state
with the largest amount of entanglement is not
the state with the smallest number of photons.
The three optimal states within the three groups
are the same three states singled out above, (\ref{op}), (\ref{minp}), and
(\ref{maxp}), and correpond to the limit $\alpha\rightarrow 0$.
\begin{figure}
\includegraphics[scale=0.4]{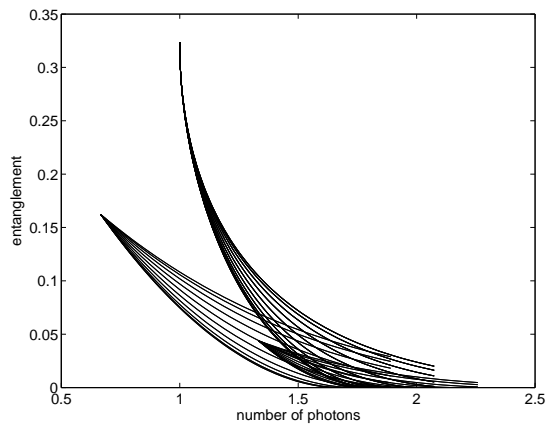}
\caption{Entanglement of formation as a function of the
average number of
photons for three types of
entangled coherent states (\ref{ech1}), (\ref{ech2})
and (\ref{ech3}). 
}
\end{figure}
We can infer two conclusions from the results plotted
in Figs.~1 and 2 that make it seem unlikely the two-mode squeezed
state would be the most robust. First, states with fewer photons are not
more robust, and second, the symmetric states of the two modes are 
not the most robust either.
 
Nevertheless, if we consider the two-mode squeezed state
with the same amount of entanglement as the other states discussed
in this subsection, then we find, using the formula
for the entanglement of symmetric Gaussian states from \cite{giedke}, 
the average number of photons in the state and the entanglement left
after decoherence are, respectively, 
$\bar{n}\approx0.5138$ and $E\approx 0.3979$, thus clearly improving on
the most robust states considered above.
 Unfortunately, without being able to calculate the 
entanglement of formation for more complicated states or states with
fewer symmetry properties or living in larger Hilbert spaces, it is not
possible to conclude anything yet about the two-mode squeezed state 
being the most robust entangled state of light in this context.
That is the main reason to consider a measure of entanglement
that can be calculated easily for any type
of entangled state, the negativity \cite{vidal1, vidal2}.
\subsection{Negativity}
Suppose we have an arbitrary pure state of two modes
$A$ and $B$. We can always write this state
in its Schmidt decomposition,
\begin{equation}
|\Psi\rangle_{AB}=\sum_k c_k |\phi_k\rangle_A|\varphi_k\rangle_B,
\end{equation}
with the $c_k$ real positive coefficients with $\sum c_k^2=1$,
and $|\phi_k\rangle_A$ and $|\varphi_k\rangle_B$
orthonormal states on systems $A$ and $B$.
The negativity of a pure state
can be defined in terms
of the Schmidt coefficients as
\begin{equation}\label{sm}
N=\frac{1}{2}\big([\sum_k c_k]^2-1\big).
\end{equation}
For a mixed state 
the negativity is determined by the sum of the absolute values
of the negative eigenvalues of the partial transpose of the density 
matrix \cite{vidal2}. Numerically, we expand each density matrix in 
number states and then the partial transpose (with respect to system $A$)
is defined as
\begin{equation}
\langle n_A,n_B|\rho^{T_A}|m_A,m_B\rangle=
\langle m_A,n_B|\rho|n_A,m_B\rangle,
\end{equation}
and the negativity is then
\begin{equation}
N(\rho)=\frac{||\rho^{T_A}||-1}{2},
\end{equation}
with $||.||$ denoting the trace norm.
We are interested in fixing the entanglement of a pure state and then
calculating the negativity of the
decohered state.
Since we have now considered two different measures of entanglement,
we will in fact always fix both the entanglement of formation and 
the negativity of the pure states. This is done as follows:
suppose we are looking for a state with $M$ Schmidt coefficients $c_1\ldots c_M$
with $M\geq 4$ and a fixed value $E$ for the entanglement of formation
and a fixed value $N$ for the negativity.
If we treat $c_3\ldots c_M$ as fixed, then $c_1$ and $c_2$
are determined by the normalization and $N$.
Since both the negativity and the norm are simple 
functions
of the Schmidt coefficients 
we can in fact determine $c_1$ and $c_2$ analytically.
The entanglement of 
formation is not a simple function, and so we use a numerical
method to determine coefficients $c_1\ldots c_3$ given coefficients $c_4\ldots c_M$. 
Namely, we use Newton's method to obtain better and better estimates of $c_3$,
such that the entanglement of formation approaches $E$,
where for each $c_3$ the coefficients $c_1$ and $c_2$ are fixed analytically.

We first consider the most straightforward case, 
where we in fact keep all Schmidt coefficients the same
as for a two-mode squeezed state, but vary the basis vectors
$|\phi_k\rangle_A$ and $|\varphi_k\rangle_B$ in the Schmidt decomposition (\ref{sm}).
An arbitrary $M$-dimensional basis can be fully
be specified by $M-1$ angles, in analogy to the 2 Euler angles in 3-D space. 
We choose the definitions of the angles such
that the number state basis $|0\rangle,|1\rangle,\ldots |M-1\rangle$
corresponds to setting all angles equal to zero. 
Instead of varying those basis vectors over all possible
choices by varying all angles over $2\pi$ independently, 
we instead choose, for computational efficiency reasons, a large number of
randomly chosen basis vectors by choosing randomly $M-1$ angles. 
In order to choose random states that
are within a certain ``distance'' from the standard Fock basis
we simply restrict the maximum values of the $M-1$ angles. 
By increasing the distance one 
can get further and further away from the
two-mode squeezed state. In Figs~3 and 4 we plot the results.
We plot the negativity of the decohered state as a function of the average photon number
in the pure state
for two different distances. The first restricts the angles to be less than
1/10, in Fig.~4 the distance is unrestricted.
The two-mode squeezed state we take has the same entanglement 
of formation as before, and $\eta=0.5$, also as before.
The Schmidt coefficients of the two-mode squeezed state become
exponentially smaller with the number of photons $n$ (see (\ref{tm}). Hence, if
we truncate the Hilbert space at a certain maximum photon number $N_m$,
the state thus obtained is a good approximation to a two-mode squeezed state provided
$N_m$ is sufficiently large.
Here we chose $N_m=6$. This means that by varying the basis states,
the maximum possible number of photons is, of course, at most 6,
as indeed is clear in Fig.~4. 
\begin{figure}
\includegraphics[scale=0.4]{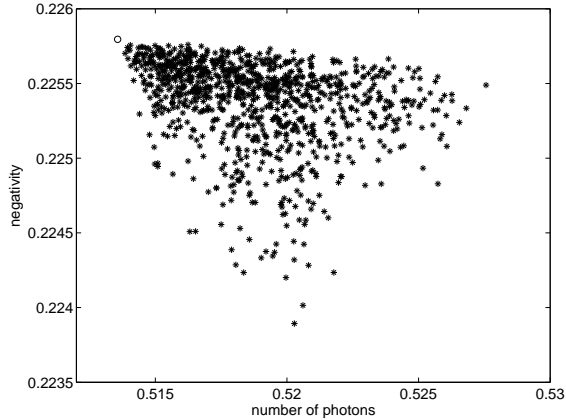}
\caption{Negativity as a function of average number of
photons for 1000 randomly generated states whose Schmidt
coefficients are equal to that of a two-mode squeezed state,
with a restriction on the distance from a two-mode squeezed state, 
as explained in the text. The circle indicates the
two-mode squeezed state.
}
\end{figure}
\begin{figure}
\includegraphics[scale=0.4]{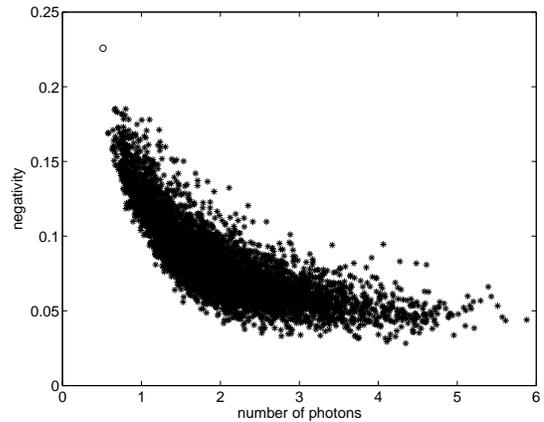}
\caption{Same as previous Figure, except there is
no restriction on the distance to the two-mode squeezed state.
}
\end{figure}
What we see from
 the two figures is that indeed the two-mode squeezed state itself is
the most robust among the set of states considered here,
which can be confirmed numerically by decreasing the distance even more
than in Fig.~3.

It is probably worthwhile noting that all entanglement properties
of a given state
are determined by the Schmidt coeffcients. And so the conclusion is
that, with all entanglement properties being equal, the state with the smallest
number of photons is the most robust,
provided we use the negativity of the decohered state to measure robustness.

As an aside we note it might be slightly confusing to 
read in Ref.~\cite{vidal1} that the negativity of any state, mixed or pure,
is proportional to
the ``robustness'' of entanglement. However, the two definitions
of the word ``robustness'' are quite different.
In \cite{vidal1} the robustness of the entanglement of a state 
refers to how much of unentangled states
on the $M$-dimensional Hlbert space has to be mixed with the original state
in order to
remove all entanglement from the state.

We can try to confirm the behavior of the negativity
for the same set of states
that featured in Fig.~1, i.e., the states with at most one photon
in each mode $A$ and $B$.
\begin{figure}
\includegraphics[scale=0.4]{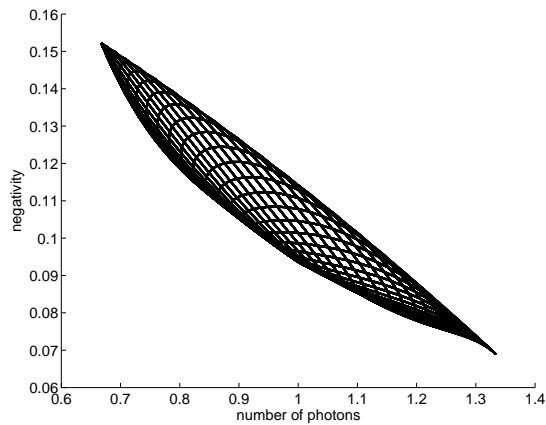}
\caption{Negativity as a function of the average number of photons
for states of the form (\ref{form1}).
}
\end{figure}
One sees that the negativity, in contrast to the entanglement of 
formation (See Fig.~1), 
does seem to favor states with fewer photons
as far as robustness is concerned. 

On the other hand, there are states
that do have the same amount of entanglement of formation
and the same negativity
as the two-mode squeeze state, but do not have the same Schmidt
coefficients. For those states there necessarily exist at least one
measure of 
entanglement that is different than that for the two-mode sqeezed state.
Let us investigate this a little further.
If we restrict the Hilbert space of both modes $A$ and $B$ to contain at 
most three photons, then there are at most four Schmidt coefficients.
The coefficients are then fixed by the normalization, by the negativity,
 by the entanglement of formation, and by some third measure of entanglement,
i.e. some other independent function of the Schmidt coefficients $c_k$.
For instance, let us, quite arbitrarily, choose the purity
of the reduced density matrix as the third entanglement measure,
\begin{eqnarray}
P=1-\sum(c_k^4).
\end{eqnarray}
We then take a two-mode squeezed state but truncate its
Fock-state expansion after four terms. In order
for it to be a good approximation to an actual two-mode squeezed
state we choose the average number of photons and the entanglement
smaller than we did in previous examples.
\begin{figure}
\includegraphics[scale=0.4]{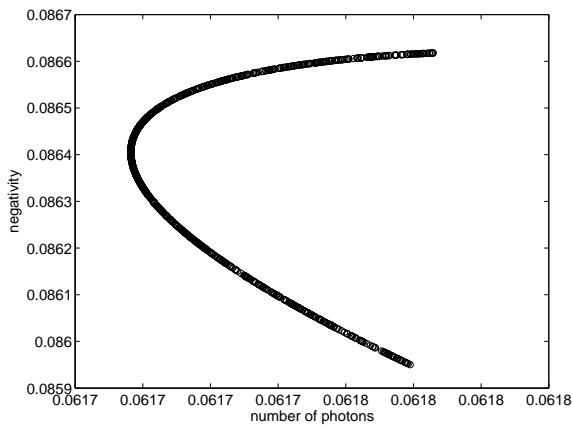}
\caption{Negativity as a function of the average number of photons
for states with fixed pure-state negativity and entanglement of formation
and at most containing three photons per mode.
}
\end{figure}
In Fig.~6 we plot the negativity of the decohered states (again using 
$\eta=0.5$) as a function of the average number of photons for 1000 randomly
generated states
that all have the same pure-state entanglement $E=0.2$ 
and negativity $N=0.2079$. The plot then shows that there
is indeed a line of points, indicating there is
exactly one  more function characterizing the initial pure state.
\begin{figure}
\includegraphics[scale=0.4]{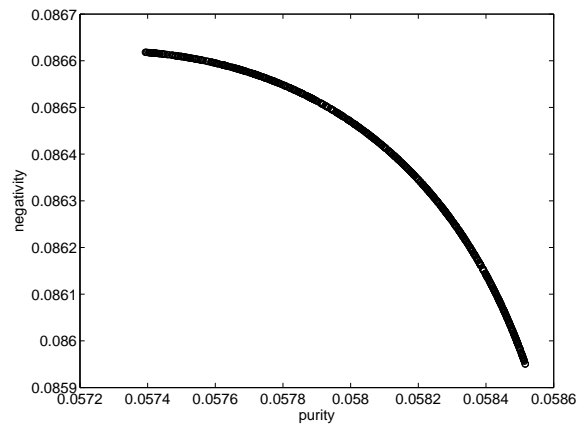}
\caption{Negativity as a function of the purity
for states with fixed pure-state negativity and entanglement of formation
and at most containing three photons per mode.
}
\end{figure}
In Fig.~7, then, we plot
the same data but as a function of the purity $P$,
which then indeed shows a monotonic behavior, rather than the
bistable behavior of Fig.~7.
In fact, it shows, perhaps surprisingly, 
that the {\em smaller} the purity, the more robust the state.
From Figs~ 6 and 7 it follows already that the two-mode squeezed state is {\em not}
the most robust entangled state of light, although it is close
in the sense that the entanglement of the most robust state is only
a trifle larger than that of the two-mode squeezed state.

We extend this discussion to states confined to rank five,
containing at most four photons.
The corresponding plots are in Figs.~8 and 9.
\begin{figure}
\includegraphics[scale=0.4]{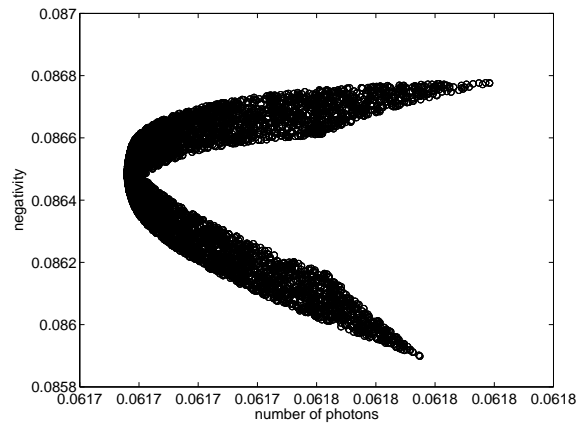}
\caption{Same as Fig.~6 except for states with at most four photons.
}
\end{figure}
\begin{figure}
\includegraphics[scale=0.4]{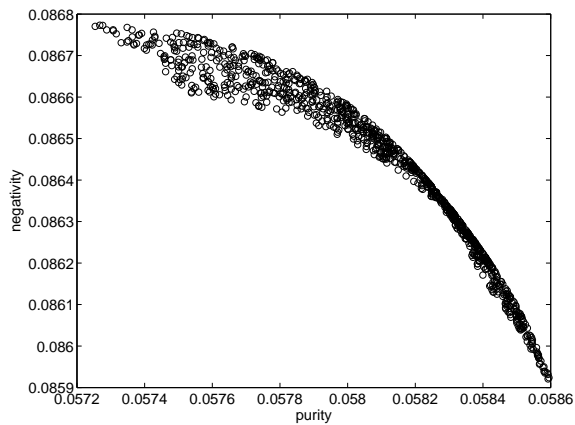}
\caption{Same as Fig.~7 except for states with at most four photons.
}
\end{figure}
Those plots show that, as expected, there is now more than one 
extra degree of freedom (in this case exactly two degrees of freedom, of course) 
that determines how robust a given entangled state is.
But the extra degrees of freedom, in this example, hardly increase (less than
0.1\%)
the negativity of the decohered state relative to that of the two-mode squeezed state.
This is because the states we considered have only a small projection
onto the extra  dimensions (i.e., the Fock states with 3 and 4 photons). 
If we increase the amount of entanglement
and, concomittantly, the allowed average number 
of photons in the initial state, 
the extra degrees of freedom
should become more important. For example, taking
states with an entanglement of $E\approx 0.82$ 
leads to an average number of photons of $\bar{n}\approx
0.5137$ for the two-mode squeezed state, and the most robust state
then turns out to be 2\% more robust than the two-mode squeezed state.
The optimum state has in that case an average photon number of 
 $\bar{n}\approx
0.5797$, as was found numerically.
Thus, we expect the two-mode squeezed state (\ref{tm}) to be close to the
most robust entangled state of light if its entanglement and 
average number of photons are small. For reasonable (i.e., experimentally achievable)
values the robustness is always clearly within 1\% of the optimum.
\section{Discussion}
In this paper we considered the degradation of entanglement
of entangled states of light suffering from
photon absorption losses. The aim was to find the most robust entangled
state, i.e., the state that preserves its entanglement best under a given
amount of noise. We found that the answer depends on how one formulates the problem.
In particular, it depends on what measure(s) of 
entanglement one uses. 
That is, we may fix any number $n$ of entanglement measures $M_1\ldots M_n$
of the initial
pure state and use one particular
entanglement measure $M_0$ of the decohered state (after photon absorption)
to define the robustness of the state. 
The answer which state is the most robust
then depends on the measures $M_0$ and $M_1\ldots M_n$ one uses.

If one fixes two measures of entanglement and takes
$M_1$ to be the entanglement of formation and $M_2$ the negativity,
and one uses the negativity for $M_0$, we found that states
with smaller number of photons tend to be more robust.
Nevertheless, the most robust state is not the state 
with the smallest possible number 
of photons,
the two-mode squeezed state.
The reason is that even when one fixes both 
the entanglement of formation and the negativity there are still
other degrees of freedom left that determine the entanglement properties of 
the pure state,
including its robustness against photon absorption losses. 
The two-mode squeezed state
is close to the most robust state, though, and is very close for 
experimentally achievable parameters. That is, the value of $M_0$
for the most robust state differs less than 1\% from that of the two-mode
squeezed state for realistic parameters.

On the other hand, if one fixes {\rm all} entanglement degrees of freedom
of a pure state (i.e. all Schmidt coefficients)
and uses the negativity of the decohered state
to quantify robustness,
then the most robust state is the one with the smallest 
(given all constraints) number of photons.


\begin{thebibliography}{99}
\bibitem{haroche}J.M.~Raimond, M.~Brune, and S.~Haroche, Phys. Rev.
Lett. {\bf 79}, 1964 (1997).

\bibitem{myatt}C.J.~Myatt {\em et al.}, Nature {\bf 403}, 269 (2000).

\bibitem{hirota}S.J.~van Enk and O.~Hirota, Phys. Rev. A {\bf 64}, 
022313 (2001).

\bibitem{lixu}S-B~Li and J-B~Xu, Phys. Lett. A {\bf 309}, 321 (2003).

\bibitem{pz}P.~Zanardi, Phys. Rev. A {\bf 65}, 042101 (2002);
S.J.~van Enk, Phys. Rev. A {\bf 67}, 022303 (2003);
Y.~Shi, Phys. Rev. A {\bf 67}, 024301 (2003).

\bibitem{wootters}W.K.~Wootters, Phys. Rev. Lett. {\bf 80}, 2245 (1998).


\bibitem{giedke}G.~Giedke {\em et al.}, Phys. Rev. Lett. {\bf 91}, 107901 
(2003).

\bibitem{vidal1}G.~Vidal and R.~Tarrach, Phys. Rev. A {\bf 59}, 141 (1999).

\bibitem{vidal2}G.~Vidal and R.F.~Werner, Phys. Rev. A {\bf 65}, 032314 (2002).


\end{thebibliography}
\end{document}